\documentclass[journal]{IEEEtran}
\usepackage{cite}
\ifCLASSINFOpdf
\else
\fi
\usepackage{cite}
\usepackage{amsmath,amssymb,amsfonts}
\usepackage{algorithmic}
\usepackage{textcomp}
\usepackage{multirow}
\usepackage{flushend}
\usepackage{graphicx}
\usepackage{colortbl}
\usepackage{caption}
\usepackage{capt-of}
\usepackage{balance}
\usepackage{subfigure}
\usepackage{fixltx2e}
\usepackage{setspace}
\usepackage{hyperref}
\usepackage{array}

\begin{document}
\title{Subject-Independent 3D Hand Kinematics Reconstruction using Pre-Movement EEG Signals for Grasp And Lift Task}
\author{Anant Jain and Lalan Kumar
\thanks{This work was supported in part by DRDO - JATC project with project number RP04191G.}
\thanks{Anant Jain is with the Department of Electrical Engineering, Indian Institute of Technology Delhi, New Delhi 110016, India (e-mail: anantjain@ee.iitd.ac.in).}
\thanks{Lalan Kumar is with the Department of Electrical Engineering, Bharti School of Telecommunication, and Yardi School of Artificial Intelligence, Indian Institute of Technology Delhi, New Delhi 110016, India (e-mail: lkumar@ee.iitd.ac.in).}
}

\markboth{Submitted to Elsevier Biomedical Signal Processing and Control}{}

\maketitle

\begin{abstract}
Brain-computer interface (BCI) systems can be utilized for kinematics decoding from scalp brain activation to control rehabilitation or power-augmenting devices. In this study, the hand kinematics decoding for grasp and lift task is performed in three-dimensional (3D) space using scalp electroencephalogram (EEG) signals. Twelve subjects from the publicly available database WAY-EEG-GAL, has been utilized in this study. In particular, multi-layer perceptron (MLP) and convolutional neural network-long short-term memory (CNN-LSTM) based deep learning frameworks are proposed that utilize the motor-neural information encoded in the pre-movement EEG data. Spectral features are analyzed for hand kinematics decoding using EEG data filtered in seven frequency ranges. The best performing frequency band spectral features has been considered for further analysis with different EEG window sizes and lag windows. Appropriate lag windows from movement onset, make the approach pre-movement in true sense. Additionally, inter-subject hand trajectory decoding analysis is performed using leave-one-subject-out (LOSO) approach. The Pearson correlation coefficient and hand trajectory are considered as performance metric to evaluate decoding performance for the neural decoders. This study explores the feasibility of inter-subject 3-D hand trajectory decoding using EEG signals only during reach and grasp task, probably for the first time. The results may provide the viable information to decode 3D hand kinematics using pre-movement EEG signals for practical BCI applications such as exoskeleton/exosuit and prosthesis.
\end{abstract}


\section{Introduction}\label{sec:introduction}

\IEEEPARstart Brain$-$computer interface (BCI) or brain$-$machine interface (BMI) utilizes brain activation for controlling external neural devices without embracing the peripheral nerves and muscles\cite{kubler2020history}. BCI is an emerging technology that demonstrate encouraging potential to ameliorate the quality of life for patients with motor impairments\cite{chaudhary2016brain,coscia2019neurotechnology,hobbs2020review} and to interact with the healthy subjects\cite{chen2019age,jin2020developing}. With the advancement in neuroscience and machine learning algorithms, BCI systems have been utilized to assist, augment, or restore the brain's motor functionality\cite{robinson2021emerging}. The BCI system is confected of subsequent processes which commonly consist of neural signal acquisition, signal processing, feature extraction, human intention detection and user feedback signal generation. The neural activity can be recorded either by invasive or non-invasive recording systems. Although the invasive approach can result in more accurate and precise brain activity recognition, the sensors are surgically placed under the scalp for acquiring the neural signals\cite{waldert2016invasive}. However, non-invasive BCI utilizes neural activity by placing sensors over the scalp. Various non-invasive techniques utilized in BCI system include magnetoencephalography (MEG)\cite{hamalainen1993meg}, electroencephalogram (EEG)\cite{niedermeyer2005eeg}, functional near-infrared spectroscopy (fNIRS)\cite{ferrari2012brief} and functional magnetic resonance imaging (fMRI)\cite{logothetis2001neurophysiological}.
\begin{figure*}[ht]
	\centering
	\includegraphics[width=0.8\textwidth]{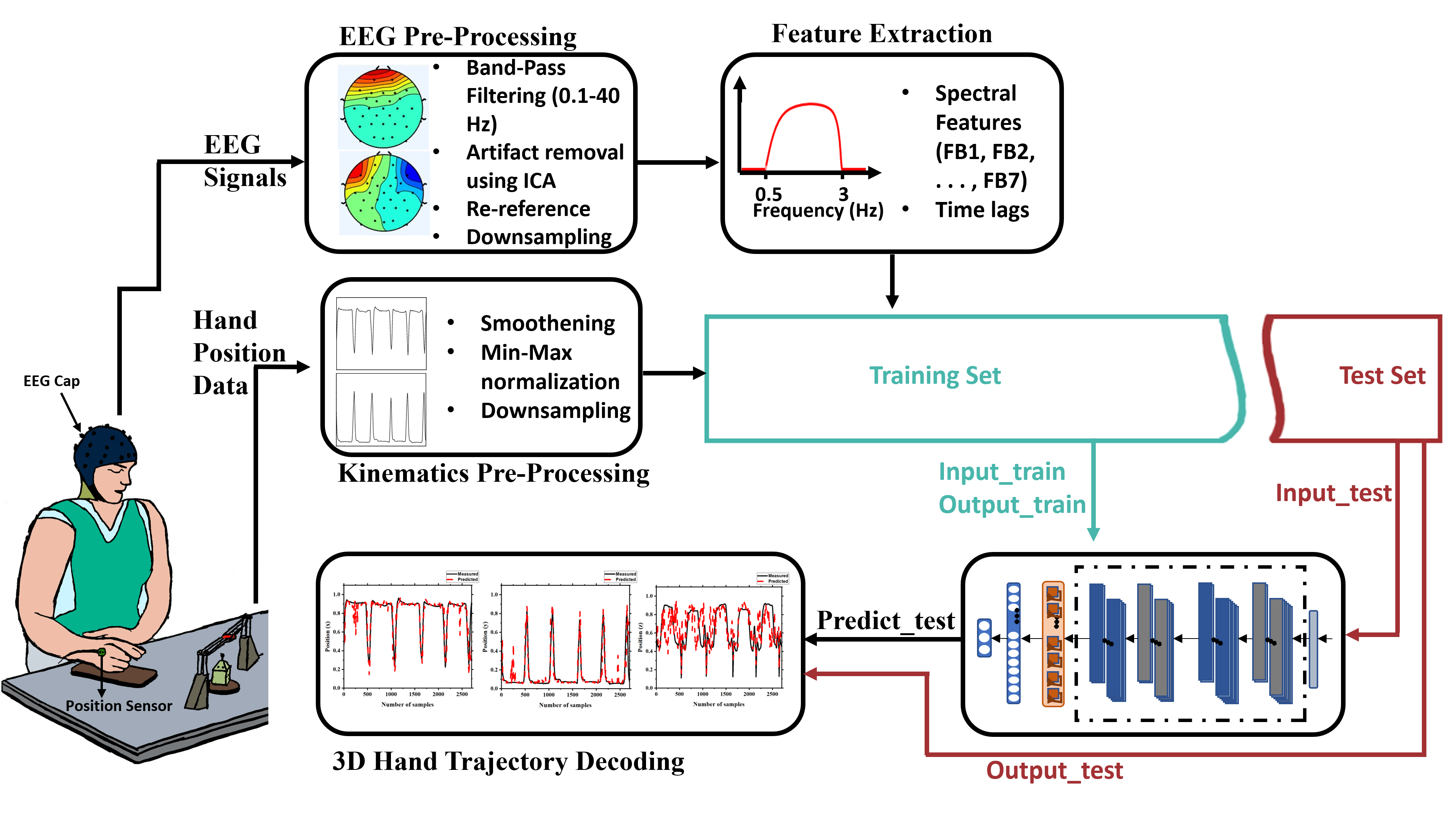}
	\caption{Flowchart of EEG-based 3-D hand kinematics decoding framework for Grasp-and-Lift task.}
	\label{exp} 
\end{figure*}

EEG based BCI system has become popular because of high temporal resolution, portability and low-economy\cite{mcfarland2017eeg}. It has been utilized for various application such as emotion recognition\cite{jana2022capsule, rajpoot2022subject}, wearable exoskeleton\cite{barsotti2015full,yu2017eeg,tariq2018eeg,al2018eeg}, and prosthesis\cite{huang2019eeg, zhu2020hybrid, nann2020restoring, cao2021brain}. It has been additionally utilized for motor imagery/execution classification\cite{ofner2017upper,chaisaen2020decoding,rammy2021sequence, rithwik2022high}, controlling robotic devices\cite{mishchenko2018developing}, and continuous classification based prosthesis control\cite{gao2019eeg}. Although classification based approach has been employed substantially, continuous kinematic estimation-based approach would yield enhanced performance and natural control of external devices such as neural exosuit, exoskeleton, or prosthesis.

Multichannel EEG-based kinematics decoding for 2-D hand trajectory was proposed in \cite{robinson2015adaptive} using multi-variable linear regression (mLR) neural decoder. In particular, a Kalman filter based mLR decoder was employed for hand trajectory decoding and the mean correlation value of $0.60\pm 0.07$ between the measured and predicted trajectory was reported. Decoding of 3-D hand trajectory using band-power EEG features and mLR as neural decoder was studied in \cite{korik2018decoding}. Using scalp EEG signals, reconstruction of hand, elbow and shoulder trajectories in 3-D space was reported in \cite{Sosnik2020} with mLR model as neural decoder. The mean Pearson's correlation coefficient between predicted and measured trajectories for hand, elbow and shoulder ranged in 0.24$-$0.49, 0.41$-$0.48 and 0.18$-$0.40, respectively. It is to be noted that mLR decoder is based on the linear mapping between input variables (EEG channels) and output variables (kinematics parameters). It is sensitive to data quality and outliers. The inadequacy of the mLR decoder can be overcome by using deep learning-based neural decoders. Deep learning decoders extract non-linear features from the input variables and are robust to handle the outliers. Deep learning-based decoders are utilized for kinematics parameter decoding\cite{jeong2020brain,source2022,jain2022premovnet} using scalp EEG signals. In \cite{jeong2020brain}, convolutional neural network - bidirectional long short-term memory (CNN-biLSTM)-based deep learning architecture was utilized for arm trajectory decoding from EEG signals to control robotic arm. The mean correlation coefficient of 0.47 was reported for movement execution. 3-D hand trajectory was decoded in \cite{source2022} for grasp and lift task using wavelet based time-frequency features and CNN-LSTM based neural decoder. Visual stimulus was considered as reference point therein for EEG signal. However, such visual stimulation limits the BCI system for day-to-day applications. A deep learning based neural decoder PreMovNet, that utilized movement onset as reference point, was introduced in \cite{jain2022premovnet} for 3D hand trajectory estimation. To the best of authors' knowledge, subject-independent analysis has not been performed for trajectory estimation during grasp and lift task. Additionally, PreMovNet, as presented in \cite{jain2022premovnet}, makes use of EEG data upto movement onset. This restricts the application of PreMovNet in EEG controlled BCI system, where controller requires advance information of trajectory for positive augmentation such as soft exosuit$/$exoskeleton. 

In this study, spectral features are analyzed for efficient hand kinematics decoding for grasp-and-lift task. The best performing spectral features are taken for further analysis to decode hand trajectory using EEG signals prior to movement onset with variable window size and distinct lag windows. Leave-one-subject-out (LOSO) technique is employed for this purpose. As motor-related neural information is encrypted in the EEG signals about 300 ms preceding the movement execution\cite{source2022}, it is suitable to use EEG signals along with appropriate lags from the motion onset, making it pre-movement based model in true sense. The organization of the article is as follows. The experimental data acquisition and data pre-processing steps are included in section \ref{sec:02_exp_setup}. The description of spectral features, neural decoders, training and evaluation of the decoding models is presented in section \ref{sec:method}. Performance evaluation for hand kinematics decoding is reported in section \ref{sec:performa}. Section \ref{sec:discussion} provides a extensive discussion of the results and section \ref{sec:conclusion} includes the conclusions about the research work.

\section{Experimental Setup}\label{sec:02_exp_setup}

\subsection{Data Acquisition}
In this analysis, open source WAY-EEG-GAL (Wearable interfaces for hAnd function recoverY- EEG - grasp and lift) database \cite{luciw2014multi} is utilized for hand kinematics decoding. 32-channel EEG recordings were collected in synchronization with hand movement. In particular, the database included recordings of twelve healthy subjects for right hand grasp and lift movement. In each trial, the subject's task to be executed was to reach and grasp a small object, lift it few centimeters up and hold it steady for a couple of seconds. The trial ended with replace and release the object to its original position, then bring the hand back to the resting position. The movement onset to reach for grasping the object and putting it down was cued by an LED. Each participant executed 294 trials of grasp and lift task with variation in object's weight (165, 330, or 660 g), contact surface (sandpaper, silk, or suede), or both.  

A LED mounted setup was put above the object, which could be switched on and off. Each trial commenced with turning on of the LED. The participant reached, grasped, lifted and held the object until the LED was activated. The LED was deactivated automatically after a couple of seconds, and then the participant put down and placed the object to its initial position. The trial ended with the participant bringing the arm back to its resting place. 

\subsection{Data Pre-processing}

Scalp EEG data was initially filtered in the frequency band of 0.1$-$40 Hz using FIR filter to expunge baseline drifts. Re-referencing of the EEG data was performed using average re-referencing method. Further, eye movement artifacts and muscle artifacts were eradicated by utilizing the independent component analysis (ICA) technique. The denoised EEG data was down-sampled to the sampling frequency of 100 Hz for reduction of computation cost. The EEG signals preprocessing was executed using the EEGLAB toolbox\cite{delorme2004eeglab}. A total of 21 EEG channels from motor cortex region ('F3', 'Fz', 'F4', 'FC5', 'FC1', 'FC2', 'FC6', 'C3', 'Cz', 'C4', 'CP5', 'CP1', 'CP2', 'CP6', 'P7', 'P3', 'Pz','P4') and occipital lobe ('O1', 'Oz', 'O2') were utilized for the hand kinematics decoding. Each of the EEG channel data was normalized using z-score normalization, given as
\begin{align}
E^{n}[t]=\frac{e^{n}[t]-\mu_{e_n}}{\sigma_{e_n}}
\label{std}
\end{align}
where, $e^{n}[t]$ and $E^{n}[t]$ are the $n^{th}$ channel pre-processed and standardized voltage, respectively, at time $t$. The mean and standard deviation of $e^{n}$ are denoted by $\mu_{e_n}$ and $\sigma_{e_n}$, respectively.

The hand kinematics data was filtered using low-pass FIR filter with a 2 Hz cut-off frequency for smoothing. Further, the min-max normalization was performed on the filtered kinematics data. Lastly, the normalized data was down-sampled to 100 Hz to match sampling frequency of EEG data.

\section{Methodology}\label{sec:method}
In this section, the description of the EEG-based 3-D hand kinematics decoding framework is presented. Fig. \ref{exp} illustrates the flow chart of the proposed hand kinematics decoding framework.

\subsection{Spectral Frequency Bands}\label{sec: spec_fb}
The EEG frequency band (FB) was segregated into groups as FB1 (delta, 0.5$-$3 Hz), FB2 (theta, 4$-$8 Hz), FB3 (alpha, 9$-$12 Hz), FB4 (beta, 13$-$30 Hz), FB5 (gamma, 30$-$50 Hz), FB6 (delta and theta, 0.5$-$8 Hz), and FB7 (delta, theta and alpha, 0.5$-$12 Hz). For hand kinematics decoding, the EEG data was filtered in each band group using a FIR filter with Hamming window. All the selected 21 EEG channels were utilized for spectral analysis. The filtered EEG data with appropriate delay (pre-movement) was utilized for the hand trajectory estimation. Three neural decoders: multi-linear regression (mLR), multi-layer perceptron (MLP) and CNN$-$LSTM model are utilized for hand kinematics decoding.

\subsection{Data Preparation}
The hand kinematics data segment of each trial was selected from the movement onset of the hand until the participant put the hand back to its resting position. The cortical activation on the motor-cortex region was observed prior to the actual movement of the hand\cite{source2022}. We assimilated the neural information corresponding to the motor activity by consolidating the EEG data prior to the hand movement. Various EEG lags and window sizes are explored for hand kinematics decoding. Here, EEG segment corresponding to time window of 100 ms could be -150 to -50 ms, where 0 ms is referenced from the movement onset. For each kinematic segment, an EEG input matrix with dimension of $D\times(L*N)$ is generated where $D$ is the data segment, $L$ is the size of time lag window, and $N$ is total EEG channels selected. 

\subsection{Multi-variable Linear Regression (mLR) model}
Multi-variable linear regression (mLR) based kinematic decoding has been utilized extensively used for brain-computer interface \cite{robinson2015adaptive,korik2018decoding,Sosnik2020}. The mLR model utilizes multiple EEG inputs to decode hand position. The mLR model has input-output mapping given by:
\begin{align}
mLR_x[t] =& \alpha_x+\sum_{n=1}^{N}\sum_{l=l_{1}}^{l_{2}}\beta^{(nl)}_{x}E^n[t-l]\\
mLR_y[t] =& \alpha_y+\sum_{n=1}^{N}\sum_{l=l_{1}}^{l_{2}}\beta^{(nl)}_{y}E^n[t-l]\\
mLR_z[t] =& \alpha_z+\sum_{n=1}^{N}\sum_{l=l_{1}}^{l_{2}}\beta^{(nl)}_{z}E^n[t-l]\\ \nonumber
\end{align}
where, $mLR_x[t]$, $mLR_y[t]$, and $mLR_z[t]$ are the position of the hand in x, y, z-directions, respectively, at time $t$. $E^n[t-l]$ is the standardized EEG signal at time lag $l$. The number of time lags is in range from $l_{1}$ to $l_{2}$ with $\alpha$ and $\beta$ as the regression coefficients of the mLR model.

\subsection{Multi-Layer Perceptron (MLP) Model}
The multi-layer perceptron (MLP) based neural decoder model consists of six layers. It includes one batch normalization layer (B1), four dense layers (D1, D2, D3 and D4), and one output layer. The initial three dense layers (D1, D2 and D3) consist of 128 neurons each, and the last dense layer (D4) consists of 16 neurons. The output layer consists of three neurons which corresponds to the hand position in the x, y, and z-directions.
\begin{table*}[t]
\centering
\caption{Spectral analysis of (a) mLR, (b) MLP, and (c) CNN-LSTM neural decoders with time lags of 150 ms, 200 ms, 250 ms, 300 ms and 350 ms.}
\scalebox{0.76}{
\centering
\begin{tabular}{|c|c|ccc|ccc|ccc|ccc|ccc|}
\hline
\multirow{2}{*}{\textbf{Frequency   Band}} & \multirow{2}{*}{\textbf{Direction}} & \multicolumn{3}{c|}{\textbf{150}}                                                      & \multicolumn{3}{c|}{\textbf{200}}                                                        & \multicolumn{3}{c|}{\textbf{250}}                                                      & \multicolumn{3}{c|}{\textbf{300}}                                                      & \multicolumn{3}{c|}{\textbf{350}}                                                        \\ \cline{3-17} 
                                           &                                     & \multicolumn{1}{c|}{\textbf{(a)}} & \multicolumn{1}{c|}{\textbf{(b)}}   & \textbf{(c)} & \multicolumn{1}{c|}{\textbf{(a)}} & \multicolumn{1}{c|}{\textbf{(b)}}   & \textbf{(c)}   & \multicolumn{1}{c|}{\textbf{(a)}} & \multicolumn{1}{c|}{\textbf{(b)}} & \textbf{(c)}   & \multicolumn{1}{c|}{\textbf{(a)}} & \multicolumn{1}{c|}{\textbf{(b)}} & \textbf{(c)}   & \multicolumn{1}{c|}{\textbf{(a)}} & \multicolumn{1}{c|}{\textbf{(b)}}   & \textbf{(c)}   \\ \hline \hline
\multirow{3}{*}{FB1}                       & x                                   & \multicolumn{1}{c|}{0.507}        & \multicolumn{1}{c|}{0.734}          & 0.764        & \multicolumn{1}{c|}{0.497}        & \multicolumn{1}{c|}{0.731}          & 0.762          & \multicolumn{1}{c|}{0.501}        & \multicolumn{1}{c|}{0.753}        & \textbf{0.791} & \multicolumn{1}{c|}{0.501}        & \multicolumn{1}{c|}{0.750}        & 0.771          & \multicolumn{1}{c|}{0.505}        & \multicolumn{1}{c|}{0.761}          & 0.771          \\ \cline{2-17} 
                                           & y                                   & \multicolumn{1}{c|}{0.518}        & \multicolumn{1}{c|}{0.743}          & 0.773        & \multicolumn{1}{c|}{0.509}        & \multicolumn{1}{c|}{0.741}          & 0.772          & \multicolumn{1}{c|}{0.512}        & \multicolumn{1}{c|}{0.761}        & \textbf{0.799} & \multicolumn{1}{c|}{0.511}        & \multicolumn{1}{c|}{0.761}        & 0.776          & \multicolumn{1}{c|}{0.516}        & \multicolumn{1}{c|}{0.769}          & 0.778          \\ \cline{2-17} 
                                           & z                                   & \multicolumn{1}{c|}{0.381}        & \multicolumn{1}{c|}{0.578}          & 0.557        & \multicolumn{1}{c|}{0.375}        & \multicolumn{1}{c|}{0.591}          & 0.582          & \multicolumn{1}{c|}{0.383}        & \multicolumn{1}{c|}{0.606}        & 0.600          & \multicolumn{1}{c|}{0.372}        & \multicolumn{1}{c|}{0.616}        & 0.598          & \multicolumn{1}{c|}{0.367}        & \multicolumn{1}{c|}{\textbf{0.621}} & 0.619          \\ \hline \hline
\multirow{3}{*}{FB2}                       & x                                   & \multicolumn{1}{c|}{0.021}        & \multicolumn{1}{c|}{0.631}          & 0.625        & \multicolumn{1}{c|}{0.033}        & \multicolumn{1}{c|}{\textbf{0.647}} & 0.654          & \multicolumn{1}{c|}{0.045}        & \multicolumn{1}{c|}{0.646}        & 0.646          & \multicolumn{1}{c|}{0.069}        & \multicolumn{1}{c|}{0.626}        & 0.639          & \multicolumn{1}{c|}{0.089}        & \multicolumn{1}{c|}{0.624}          & 0.636          \\ \cline{2-17} 
                                           & y                                   & \multicolumn{1}{c|}{0.022}        & \multicolumn{1}{c|}{0.654}          & 0.649        & \multicolumn{1}{c|}{0.034}        & \multicolumn{1}{c|}{0.667}          & 0.676          & \multicolumn{1}{c|}{0.046}        & \multicolumn{1}{c|}{0.663}        & \textbf{0.668} & \multicolumn{1}{c|}{0.070}        & \multicolumn{1}{c|}{0.643}        & 0.660          & \multicolumn{1}{c|}{0.090}        & \multicolumn{1}{c|}{0.639}          & 0.656          \\ \cline{2-17} 
                                           & z                                   & \multicolumn{1}{c|}{0.025}        & \multicolumn{1}{c|}{0.412}          & 0.375        & \multicolumn{1}{c|}{0.034}        & \multicolumn{1}{c|}{\textbf{0.418}} & 0.382          & \multicolumn{1}{c|}{0.047}        & \multicolumn{1}{c|}{0.415}        & 0.377          & \multicolumn{1}{c|}{0.063}        & \multicolumn{1}{c|}{0.423}        & 0.377          & \multicolumn{1}{c|}{0.078}        & \multicolumn{1}{c|}{0.423}          & 0.400          \\ \hline \hline
\multirow{3}{*}{FB3}                       & x                                   & \multicolumn{1}{c|}{0.137}        & \multicolumn{1}{c|}{0.500}          & 0.493        & \multicolumn{1}{c|}{0.239}        & \multicolumn{1}{c|}{0.496}          & 0.498          & \multicolumn{1}{c|}{0.283}        & \multicolumn{1}{c|}{0.518}        & \textbf{0.524} & \multicolumn{1}{c|}{0.333}        & \multicolumn{1}{c|}{0.502}        & 0.499          & \multicolumn{1}{c|}{0.365}        & \multicolumn{1}{c|}{0.503}          & 0.515          \\ \cline{2-17} 
                                           & y                                   & \multicolumn{1}{c|}{0.142}        & \multicolumn{1}{c|}{0.523}          & 0.515        & \multicolumn{1}{c|}{0.246}        & \multicolumn{1}{c|}{0.514}          & 0.521          & \multicolumn{1}{c|}{0.288}        & \multicolumn{1}{c|}{0.537}        & \textbf{0.544} & \multicolumn{1}{c|}{0.340}        & \multicolumn{1}{c|}{0.520}        & 0.520          & \multicolumn{1}{c|}{0.375}        & \multicolumn{1}{c|}{0.519}          & 0.485          \\ \cline{2-17} 
                                           & z                                   & \multicolumn{1}{c|}{0.094}        & \multicolumn{1}{c|}{\textbf{0.427}} & 0.417        & \multicolumn{1}{c|}{0.172}        & \multicolumn{1}{c|}{0.414}          & 0.396          & \multicolumn{1}{c|}{0.207}        & \multicolumn{1}{c|}{0.403}        & 0.394          & \multicolumn{1}{c|}{0.213}        & \multicolumn{1}{c|}{0.396}        & 0.367          & \multicolumn{1}{c|}{0.216}        & \multicolumn{1}{c|}{0.394}          & 0.364          \\ \hline \hline
\multirow{3}{*}{FB4}                       & x                                   & \multicolumn{1}{c|}{0.129}        & \multicolumn{1}{c|}{0.399}          & 0.403        & \multicolumn{1}{c|}{0.194}        & \multicolumn{1}{c|}{0.423}          & 0.444          & \multicolumn{1}{c|}{0.227}        & \multicolumn{1}{c|}{0.427}        & 0.465          & \multicolumn{1}{c|}{0.257}        & \multicolumn{1}{c|}{0.435}        & 0.489          & \multicolumn{1}{c|}{0.278}        & \multicolumn{1}{c|}{0.431}          & \textbf{0.511} \\ \cline{2-17} 
                                           & y                                   & \multicolumn{1}{c|}{0.135}        & \multicolumn{1}{c|}{0.419}          & 0.429        & \multicolumn{1}{c|}{0.201}        & \multicolumn{1}{c|}{0.439}          & 0.472          & \multicolumn{1}{c|}{0.235}        & \multicolumn{1}{c|}{0.441}        & 0.494          & \multicolumn{1}{c|}{0.264}        & \multicolumn{1}{c|}{0.454}        & 0.515          & \multicolumn{1}{c|}{0.288}        & \multicolumn{1}{c|}{0.445}          & \textbf{0.530} \\ \cline{2-17} 
                                           & z                                   & \multicolumn{1}{c|}{0.081}        & \multicolumn{1}{c|}{0.376}          & 0.413        & \multicolumn{1}{c|}{0.123}        & \multicolumn{1}{c|}{0.375}          & \textbf{0.442} & \multicolumn{1}{c|}{0.143}        & \multicolumn{1}{c|}{0.375}        & \textbf{0.442} & \multicolumn{1}{c|}{0.153}        & \multicolumn{1}{c|}{0.376}        & 0.433          & \multicolumn{1}{c|}{0.162}        & \multicolumn{1}{c|}{0.367}          & 0.430          \\ \hline \hline
\multirow{3}{*}{FB5}                       & x                                   & \multicolumn{1}{c|}{0.090}        & \multicolumn{1}{c|}{0.275}          & 0.296        & \multicolumn{1}{c|}{0.106}        & \multicolumn{1}{c|}{0.295}          & 0.327          & \multicolumn{1}{c|}{0.123}        & \multicolumn{1}{c|}{0.274}        & 0.352          & \multicolumn{1}{c|}{0.152}        & \multicolumn{1}{c|}{0.234}        & 0.349          & \multicolumn{1}{c|}{0.178}        & \multicolumn{1}{c|}{0.263}          & \textbf{0.371} \\ \cline{2-17} 
                                           & y                                   & \multicolumn{1}{c|}{0.097}        & \multicolumn{1}{c|}{0.291}          & 0.316        & \multicolumn{1}{c|}{0.113}        & \multicolumn{1}{c|}{0.311}          & 0.352          & \multicolumn{1}{c|}{0.129}        & \multicolumn{1}{c|}{0.290}        & 0.381          & \multicolumn{1}{c|}{0.159}        & \multicolumn{1}{c|}{0.248}        & 0.376          & \multicolumn{1}{c|}{0.187}        & \multicolumn{1}{c|}{0.278}          & \textbf{0.398} \\ \cline{2-17} 
                                           & z                                   & \multicolumn{1}{c|}{0.076}        & \multicolumn{1}{c|}{0.229}          & 0.262        & \multicolumn{1}{c|}{0.090}        & \multicolumn{1}{c|}{0.253}          & 0.276          & \multicolumn{1}{c|}{0.101}        & \multicolumn{1}{c|}{0.220}        & \textbf{0.278} & \multicolumn{1}{c|}{0.117}        & \multicolumn{1}{c|}{0.189}        & 0.266          & \multicolumn{1}{c|}{0.129}        & \multicolumn{1}{c|}{0.196}          & 0.269          \\ \hline \hline
\multirow{3}{*}{FB6}                       & x                                   & \multicolumn{1}{c|}{0.460}        & \multicolumn{1}{c|}{0.720}          & 0.740        & \multicolumn{1}{c|}{0.460}        & \multicolumn{1}{c|}{0.727}          & 0.754          & \multicolumn{1}{c|}{0.466}        & \multicolumn{1}{c|}{0.748}        & 0.768          & \multicolumn{1}{c|}{0.469}        & \multicolumn{1}{c|}{0.744}        & \textbf{0.775} & \multicolumn{1}{c|}{0.476}        & \multicolumn{1}{c|}{0.744}          & 0.766          \\ \cline{2-17} 
                                           & y                                   & \multicolumn{1}{c|}{0.471}        & \multicolumn{1}{c|}{0.728}          & 0.745        & \multicolumn{1}{c|}{0.472}        & \multicolumn{1}{c|}{0.736}          & 0.762          & \multicolumn{1}{c|}{0.477}        & \multicolumn{1}{c|}{0.757}        & 0.777          & \multicolumn{1}{c|}{0.480}        & \multicolumn{1}{c|}{0.752}        & \textbf{0.779} & \multicolumn{1}{c|}{0.487}        & \multicolumn{1}{c|}{0.753}          & 0.770          \\ \cline{2-17} 
                                           & z                                   & \multicolumn{1}{c|}{0.350}        & \multicolumn{1}{c|}{0.571}          & 0.570        & \multicolumn{1}{c|}{0.355}        & \multicolumn{1}{c|}{0.586}          & 0.589          & \multicolumn{1}{c|}{0.351}        & \multicolumn{1}{c|}{0.611}        & 0.602          & \multicolumn{1}{c|}{0.344}        & \multicolumn{1}{c|}{0.606}        & 0.608          & \multicolumn{1}{c|}{0.343}        & \multicolumn{1}{c|}{\textbf{0.622}} & 0.616          \\ \hline \hline
\multirow{3}{*}{FB7}                       & x                                   & \multicolumn{1}{c|}{0.432}        & \multicolumn{1}{c|}{0.715}          & 0.723        & \multicolumn{1}{c|}{0.436}        & \multicolumn{1}{c|}{0.721}          & 0.744          & \multicolumn{1}{c|}{0.442}        & \multicolumn{1}{c|}{0.727}        & \textbf{0.766} & \multicolumn{1}{c|}{0.452}        & \multicolumn{1}{c|}{0.738}        & 0.752          & \multicolumn{1}{c|}{0.461}        & \multicolumn{1}{c|}{0.730}          & 0.753          \\ \cline{2-17} 
                                           & y                                   & \multicolumn{1}{c|}{0.443}        & \multicolumn{1}{c|}{0.722}          & 0.731        & \multicolumn{1}{c|}{0.448}        & \multicolumn{1}{c|}{0.729}          & 0.752          & \multicolumn{1}{c|}{0.452}        & \multicolumn{1}{c|}{0.737}        & \textbf{0.770} & \multicolumn{1}{c|}{0.461}        & \multicolumn{1}{c|}{0.743}        & 0.761          & \multicolumn{1}{c|}{0.472}        & \multicolumn{1}{c|}{0.737}          & 0.757          \\ \cline{2-17} 
                                           & z                                   & \multicolumn{1}{c|}{0.332}        & \multicolumn{1}{c|}{0.589}          & 0.594        & \multicolumn{1}{c|}{0.335}        & \multicolumn{1}{c|}{0.603}          & 0.618          & \multicolumn{1}{c|}{0.334}        & \multicolumn{1}{c|}{0.608}        & 0.627          & \multicolumn{1}{c|}{0.329}        & \multicolumn{1}{c|}{0.613}        & 0.630          & \multicolumn{1}{c|}{0.329}        & \multicolumn{1}{c|}{0.619}          & \textbf{0.635} \\ \hline
\end{tabular}}
\label{tab:tab03}
\vspace{0.10cm}
\scriptsize{\\
Note: the bold entries represent the highest PCC value obtained using the neural decoders in x, y and z directions for each frequency band.}
\end{table*}

\subsection{CNN-LSTM Model}
In this Section, a CNN-LSTM based neural decoder is detailed for hand kinematics decoding during grasp and lift task. The neural decoder comprises of nine layers which include a batch normalization layer, two convolution layers ($C_1$ and $C_2$), two max-pooling layers ($M_1$ and $M_2$), one dropout layer, one LSTM layer ($L_1$), and two dense layers ($D_1$ and $D_2$). The $C_1$ and $C_2$ layers have a kernel size of $7$ and $5$ with 256 and 128 filters, respectively. Zero padding is also utilized for both $C_1$ and $C_2$ layers that results in same input and output size. ReLu activation unit is used in each convolution layer. Max-pooling layers, $M_1$ and $M_2$, have window size of $5$ and $3$, respectively. Dropout layer has a dropout rate of 0.25. LSTM layer, $L_1$, consists of 128 cells, along with the ReLU activation function. Dense layers, $D_1$ and $D_2$, have 128 and 3 neurons, respectively. The output layer consist of three neurons that yields the predicted hand position in the x, y, and z-directions.

\subsection{Training and Evaluation}
For training and performance evaluation of neural decoders, the data-set is divided into distinct training, validation, and test data. The training data is utilized for training the decoders, while the validation data is used for tuning model hyper-parameters and avoiding over-fitting of the decoders. The test data is utilized to evaluate performance of the trained neural decoders. The adaptive moment estimation (Adam) optimization algorithm\cite{kingma2014adam} with loss function as mean squared error is adapted to train neural decoders based on deep learning architectures. The early stopping technique with patience of five epochs on validation data is utilized to avoid over-fitting of the decoding models. For subject-dependent analysis, a total of 294 trials are taken from each participant of the WAY-EEG-GAL data-set. The total trials for each participant are separated into three discrete subsets: (a) 234 trials data samples as training data; (b) 30 trials data samples as validation data; and (c) 30 trials data samples as test data.

For inter-subject analysis, the data from eleven participants is taken for training$-$validation and the total trials of remaining one participant are taken as testing data to evaluate the trained model. This inter-subject decoding analysis approach is known as leave-one-subject-out (LOSO) technique. The training data is further divided into two subsets: (a) training subset and (b) validation subset. Training subset consists of 264 trials from each of eleven participants' data while the validation subset include remaining 30 trials from each of the selected eleven participants. Test set includes all 294 trials of the one left out subject to evaluate the decoding performance of the model. Pearson's correlation coefficient (PCC) between the predicted hand kinematics and measured kinematics data is taken as performance metric to evaluate the decoding performance of the neural decoders. In particular, the training step is computationally rigorous, while the estimation with trained model is brisk. Therefore, the trained decoding model can be implemented to control external devices such as exoskeletons/exosuits or prosthesis.

\section{Performance Evaluation}\label{sec:performa}
Pearson correlation coefficient (PCC) is considered as performance metric to evaluate the efficiency of the neural decoders for hand kinematics decoding. In particular, PCC is computed with various time lag windows. PCC is a linear correlation coefficient whose value ranges from $-1$ to $+1$. A $-1$, $0$ and $+1$ PCC value represents a strong negative, zero and strong positive correlation respectively. Pearson correlation coefficient between measured ($P_x$) and estimated ($P_y$) hand kinematic parameters with a total samples of $T$ is defined as
\begin{align}
	\Pi (P_x,P_y)=\frac{1}{T-1}\sum_{i=1}^{T}\left ( \frac{P_x^i-\mu^{P_x}}{\sigma^{P_x}} \right )\left( \frac{P_y^i-\mu^{P_y}}{\sigma^{P_y}} \right )
	\label{corr}
\end{align}
where, $\mu_q$ and $\sigma_q$ are the mean and standard deviation of $q$, respectively, with $q\in \{P_x,P_y\}$.

 Additionally, the hand trajectory estimation in x, y, and z-directions are plotted along with the measured trajectory for comparative analysis.
\begin{figure*}[!t]
	\centering
	\subfigure[]{\includegraphics[width=0.28\textwidth]{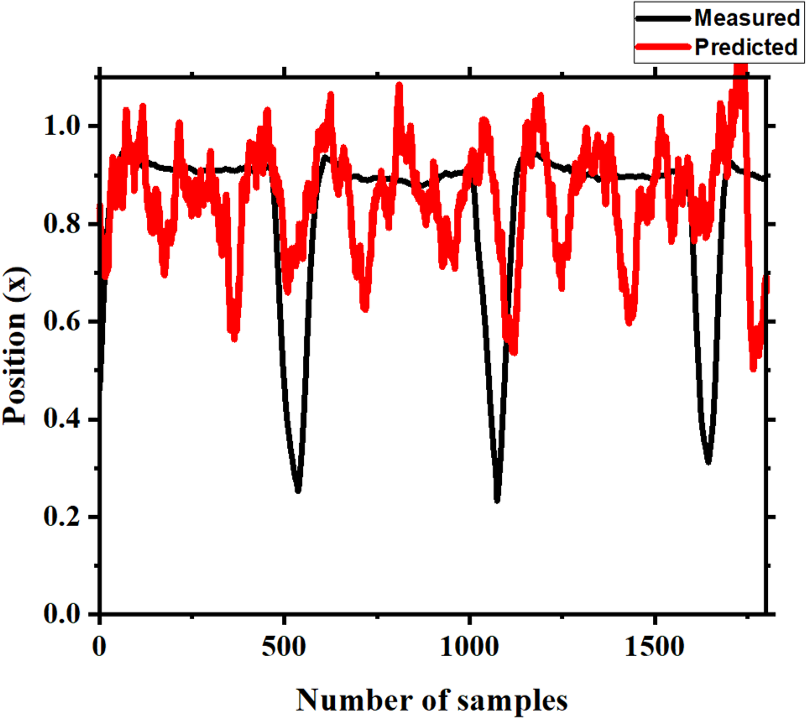}}
	\subfigure[]{\includegraphics[width=0.28\textwidth]{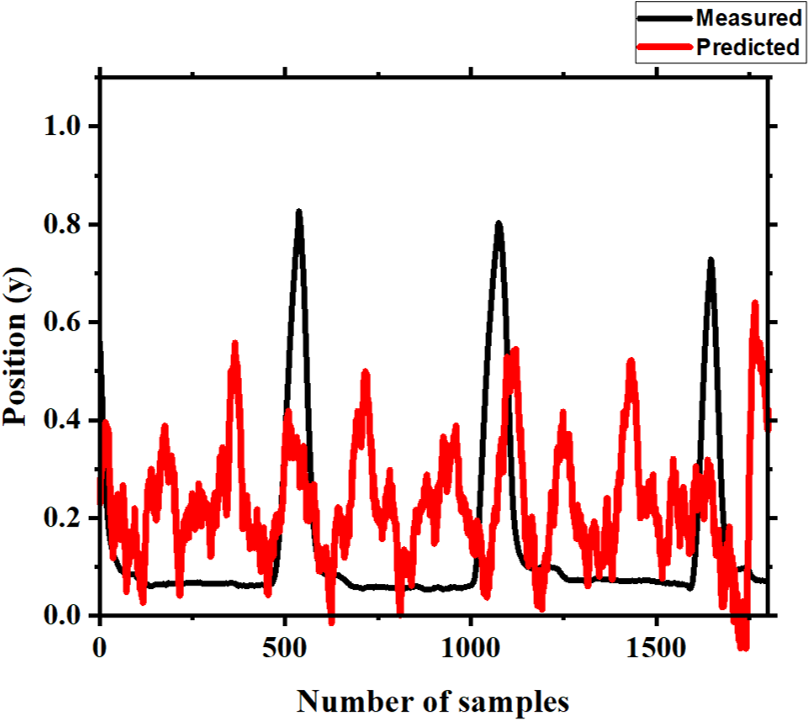}}
	\subfigure[]{\includegraphics[width=0.28\textwidth]{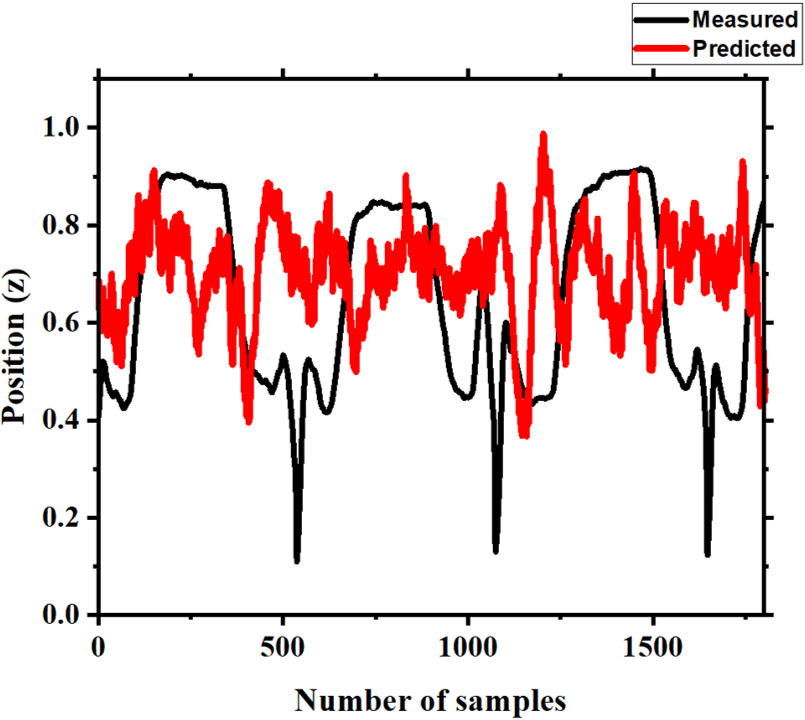}}
	\subfigure[]{\includegraphics[width=0.28\textwidth]{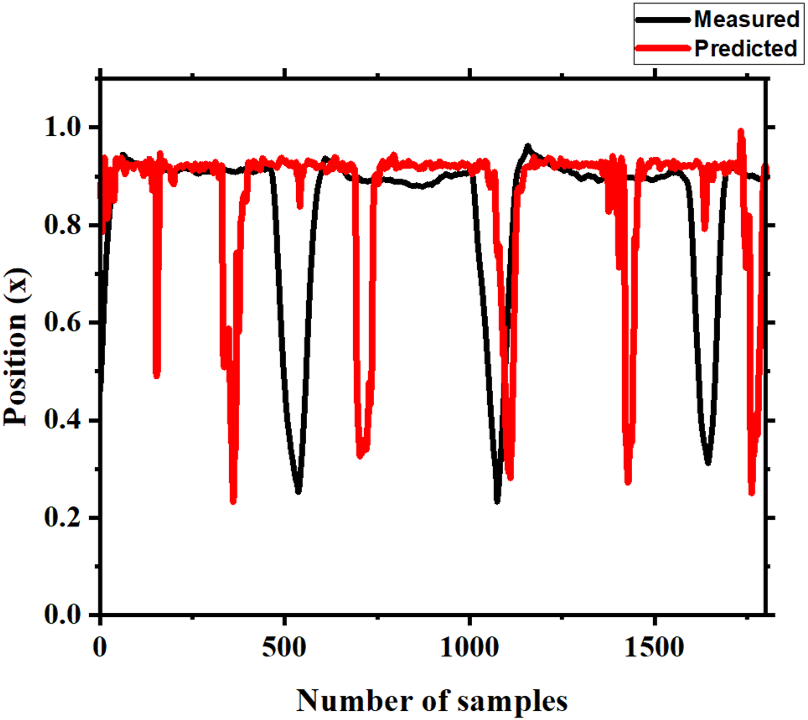}}
	\subfigure[]{\includegraphics[width=0.28\textwidth]{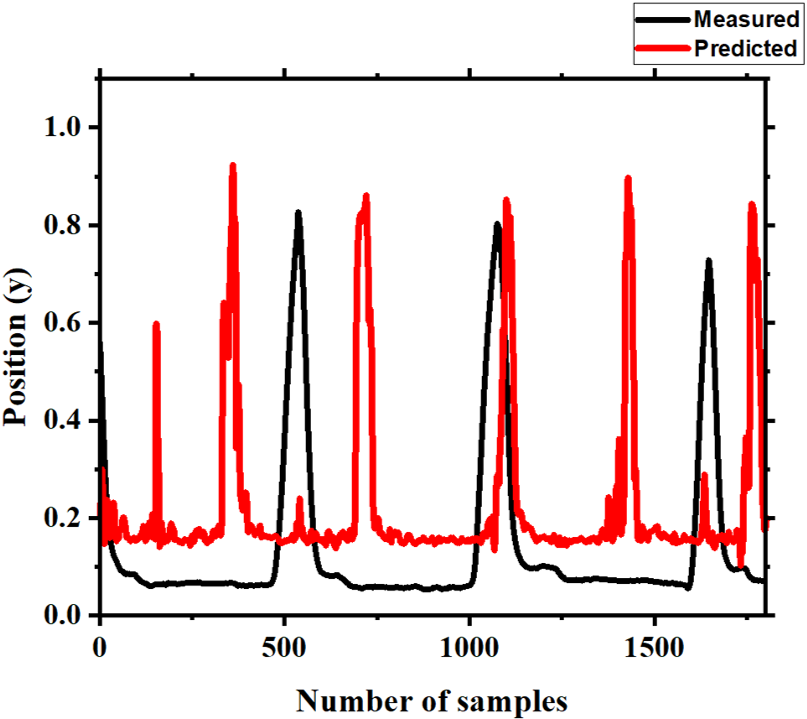}}
	\subfigure[]{\includegraphics[width=0.28\textwidth]{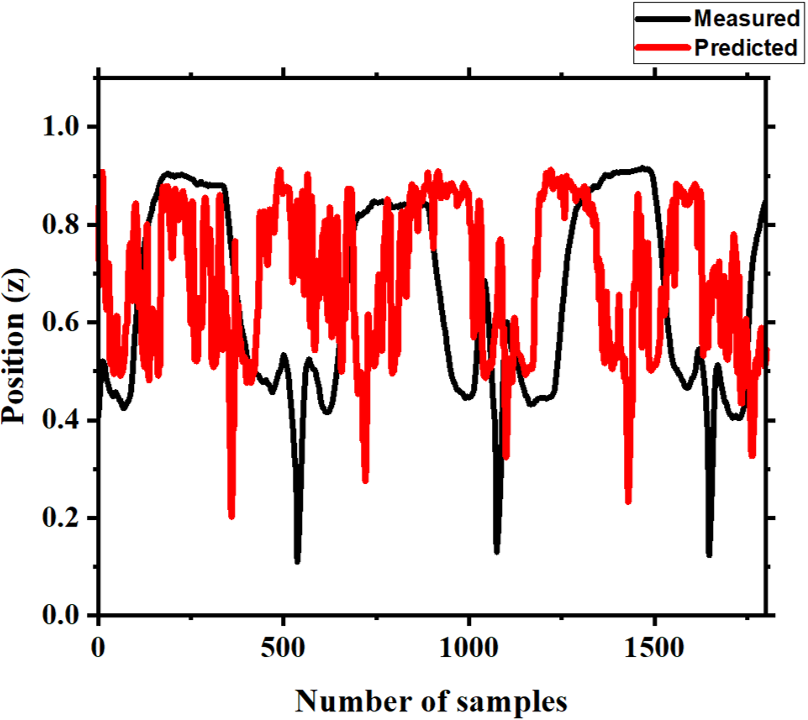}}
	\subfigure[]{\includegraphics[width=0.28\textwidth]{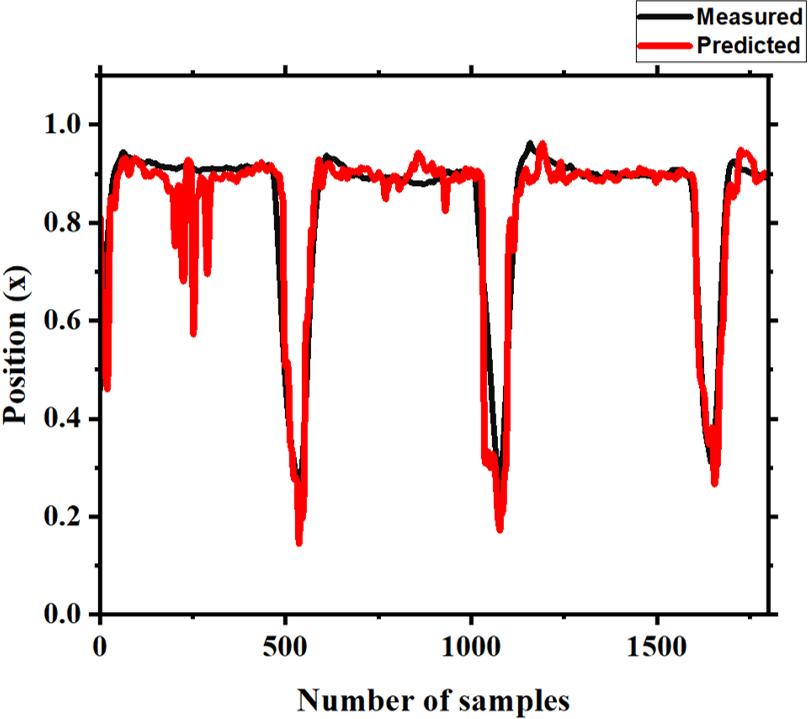}}
	\subfigure[]{\includegraphics[width=0.28\textwidth]{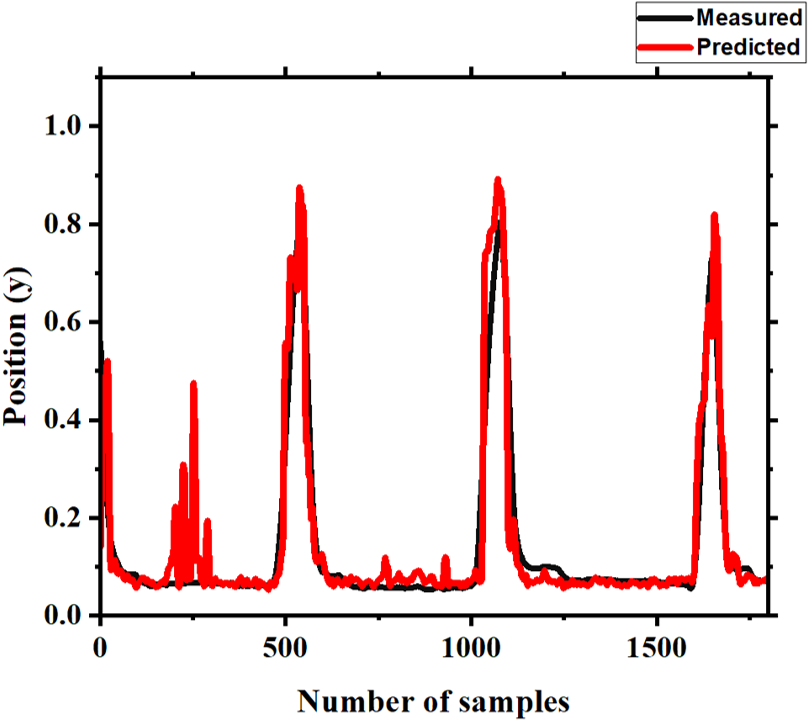}}
	\subfigure[]{\includegraphics[width=0.28\textwidth]{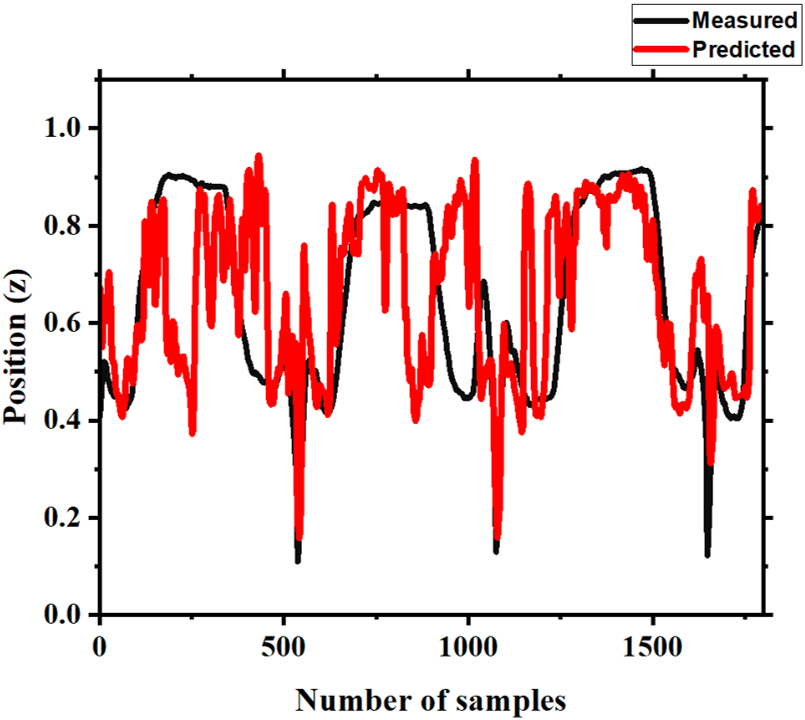}}
	\caption{Hand Trajectory Decoding in x, y and z-directions using mLR decoder is presented in (a)-(c) respectively,using MLP decoder is presented in (d)-(f) respectively, and using CNN-LSTM decoder is presented in (g)-(i) respectively.}
	\label{trajectory1}
\end{figure*} 
\subsection{Spectral Analysis}\label{sec:spectral}

The 3-D hand trajectory decoding is performed on seven FBs (as detailed in section \ref{sec: spec_fb}) with EEG lags for twelve participants from WAY-EEG-GAL database. Three neural decoders are utilized for this purpose. The mean PCC values for distinct FBs and EEG time lags are presented in Table-\ref{tab:tab03} for x, y and z-directions. In x and y-directions, the spectral features in FB1 frequency band with 250 ms EEG lag provides best PCC values for CNN-LSTM based neural decoder. However, in z-direction, CNN-LSTM based neural decoder with spectral features in FB7 frequency band and EEG lag of 350 ms gives best PCC value. It may be observed that MLP and CNN-LSTM based deep learning decoders perform significantly better than the mLR neural decoder. The lower correlation values in z-direction is due to transient movement.

\subsection{Trajectory Analysis}\label{sec: trajec_analysis}

In this section, the predicted hand trajectory is compared herein with actual trajectory in x, y and z-directions. The measured and predicted hand trajectories in x, y and z-directions are plotted in Fig. \ref{trajectory1}(a)$-$(c) for mLR neural decoder, Fig. \ref{trajectory1}(d)$-$(f) for MLP model and Fig. \ref{trajectory1}(g)-(i) for CNN-LSTM model. In Fig \ref{trajectory1}, hand trajectories are plotted for participant 04 with spectral features in FB1 frequency band and EEG lag of 200 ms. For MLP and CNN-LSTM based deep learning neural decoders, lower trajectory mismatch is observed along x,y and z-directions when compared with mLR model.

\subsection{EEG lags analysis}\label{sec:eeg_lag}
In spectral analysis, it may be observed that the decoding performance of neural decoders with FB1quency band spectral features in x and y-directions is notably better. The performance is similar with FB7 frequency band spectral features in z-direction. Therefore, spectral features in FB1 frequency band is taken for further analysis. In this analysis, the hand movement trajectory is decoded using pre-movement EEG data with distinct window sizes and lags as shown in Table \ref{tab:tab04}. It may be observed that 3D hand trajectory decoding using pre-movement EEG signals is feasible for grasp and lift task.

\begin{table}[ht]
\caption{EEG lag analysis of (a) mLR, (b) MLP, and (c) CNN-LSTM neural decoders with various lag window and window sizes.}
\scalebox{0.83}{
\centering
\begin{tabular}{|cc|c|c|c|c|}
\hline
\multicolumn{2}{|c|}{\textbf{EEG   Lags}}                                                & \multirow{2}{*}{\textbf{Decoders}} & \multirow{2}{*}{\textbf{x}} & \multirow{2}{*}{\textbf{y}} & \multirow{2}{*}{\textbf{z}} \\ \cline{1-2}
\multicolumn{1}{|l|}{\textbf{Window Size}}           & \textbf{Lag Window}               &                                    &                             &                             &                             \\ \hline
\multicolumn{1}{|c|}{\multirow{15}{*}{\textbf{100}}} & \multirow{3}{*}{\textbf{150-50}}  & \textbf{(a)}                       & 0.486                       & 0.497                       & 0.373                       \\ \cline{3-6} 
\multicolumn{1}{|c|}{}                               &                                   & \textbf{(b)}                       & 0.735                       & 0.745                       & 0.578                       \\ \cline{3-6} 
\multicolumn{1}{|c|}{}                               &                                   & \textbf{(c)}                       & 0.735                       & 0.744                       & 0.553                       \\ \cline{2-6} 
\multicolumn{1}{|c|}{}                               & \multirow{3}{*}{\textbf{200-100}} & \textbf{(a)}                       & 0.466                       & 0.478                       & 0.367                       \\ \cline{3-6} 
\multicolumn{1}{|c|}{}                               &                                   & \textbf{(b)}                       & 0.729                       & 0.737                       & 0.585                       \\ \cline{3-6} 
\multicolumn{1}{|c|}{}                               &                                   & \textbf{(c)}                       & 0.730                       & 0.734                       & 0.555                       \\ \cline{2-6} 
\multicolumn{1}{|c|}{}                               & \multirow{3}{*}{\textbf{250-150}} & \textbf{(a)}                       & 0.451                       & 0.461                       & 0.357                       \\ \cline{3-6} 
\multicolumn{1}{|c|}{}                               &                                   & \textbf{(b)}                       & \textbf{0.742}              & \textbf{0.751}              & \textbf{0.582}              \\ \cline{3-6} 
\multicolumn{1}{|c|}{}                               &                                   & \textbf{(c)}                       & 0.734                       & 0.740                       & 0.558                       \\ \cline{2-6} 
\multicolumn{1}{|c|}{}                               & \multirow{3}{*}{\textbf{300-200}} & \textbf{(a)}                       & 0.440                       & 0.451                       & 0.340                       \\ \cline{3-6} 
\multicolumn{1}{|c|}{}                               &                                   & \textbf{(b)}                       & 0.726                       & 0.734                       & 0.579                       \\ \cline{3-6} 
\multicolumn{1}{|c|}{}                               &                                   & \textbf{(c)}                       & 0.735                       & 0.741                       & 0.555                       \\ \cline{2-6} 
\multicolumn{1}{|c|}{}                               & \multirow{3}{*}{\textbf{350-250}} & \textbf{(a)}                       & 0.445                       & 0.457                       & 0.335                       \\ \cline{3-6} 
\multicolumn{1}{|c|}{}                               &                                   & \textbf{(b)}                       & 0.732                       & 0.741                       & 0.581                       \\ \cline{3-6} 
\multicolumn{1}{|c|}{}                               &                                   & \textbf{(c)}                       & 0.738                       & 0.745                       & 0.560                       \\ \hline \hline
\multicolumn{1}{|c|}{\multirow{12}{*}{\textbf{150}}} & \multirow{3}{*}{\textbf{200-50}}  & \textbf{(a)}                       & 0.489                       & 0.501                       & 0.375                       \\ \cline{3-6} 
\multicolumn{1}{|c|}{}                               &                                   & \textbf{(b)}                       & 0.742                       & 0.750                       & \textbf{0.596}              \\ \cline{3-6} 
\multicolumn{1}{|c|}{}                               &                                   & \textbf{(c)}                       & 0.732                       & 0.741                       & 0.558                       \\ \cline{2-6} 
\multicolumn{1}{|c|}{}                               & \multirow{3}{*}{\textbf{250-100}} & \textbf{(a)}                       & 0.475                       & 0.485                       & 0.364                       \\ \cline{3-6} 
\multicolumn{1}{|c|}{}                               &                                   & \textbf{(b)}                       & 0.741                       & 0.750                       & 0.579                       \\ \cline{3-6} 
\multicolumn{1}{|c|}{}                               &                                   & \textbf{(c)}                       & 0.735                       & 0.742                       & 0.548                       \\ \cline{2-6} 
\multicolumn{1}{|c|}{}                               & \multirow{3}{*}{\textbf{300-150}} & \textbf{(a)}                       & 0.464                       & 0.475                       & 0.348                       \\ \cline{3-6} 
\multicolumn{1}{|c|}{}                               &                                   & \textbf{(b)}                       & 0.744                       & 0.751                       & 0.593                       \\ \cline{3-6} 
\multicolumn{1}{|c|}{}                               &                                   & \textbf{(c)}                       & 0.741                       & 0.744                       & 0.554                       \\ \cline{2-6} 
\multicolumn{1}{|c|}{}                               & \multirow{3}{*}{\textbf{350-200}} & \textbf{(a)}                       & 0.468                       & 0.480                       & 0.342                       \\ \cline{3-6} 
\multicolumn{1}{|c|}{}                               &                                   & \textbf{(b)}                       & \textbf{0.745}              & \textbf{0.753}              & \textbf{0.596}              \\ \cline{3-6} 
\multicolumn{1}{|c|}{}                               &                                   & \textbf{(c)}                       & 0.737                       & 0.741                       & 0.560                       \\ \hline \hline
\multicolumn{1}{|c|}{\multirow{9}{*}{\textbf{200}}}  & \multirow{3}{*}{\textbf{250-50}}  & \textbf{(a)}                       & 0.490                       & 0.501                       & 0.375                       \\ \cline{3-6} 
\multicolumn{1}{|c|}{}                               &                                   & \textbf{(b)}                       & \textbf{0.751}              & \textbf{0.761}              & 0.599                       \\ \cline{3-6} 
\multicolumn{1}{|c|}{}                               &                                   & \textbf{(c)}                       & 0.749                       & 0.753                       & 0.575                       \\ \cline{2-6} 
\multicolumn{1}{|c|}{}                               & \multirow{3}{*}{\textbf{300-100}} & \textbf{(a)}                       & 0.478                       & 0.490                       & 0.359                       \\ \cline{3-6} 
\multicolumn{1}{|c|}{}                               &                                   & \textbf{(b)}                       & 0.745                       & 0.753                       & 0.604                       \\ \cline{3-6} 
\multicolumn{1}{|c|}{}                               &                                   & \textbf{(c)}                       & 0.737                       & 0.743                       & 0.575                       \\ \cline{2-6} 
\multicolumn{1}{|c|}{}                               & \multirow{3}{*}{\textbf{350-150}} & \textbf{(a)}                       & 0.480                       & 0.492                       & 0.350                       \\ \cline{3-6} 
\multicolumn{1}{|c|}{}                               &                                   & \textbf{(b)}                       & \textbf{0.751}              & 0.760                       & \textbf{0.607}              \\ \cline{3-6} 
\multicolumn{1}{|c|}{}                               &                                   & \textbf{(c)}                       & 0.749                       & 0.752                       & 0.565                       \\ \hline \hline
\multicolumn{1}{|c|}{\multirow{6}{*}{\textbf{250}}}  & \multirow{3}{*}{\textbf{300-50}}  & \textbf{(a)}                       & 0.490                       & 0.501                       & 0.367                       \\ \cline{3-6} 
\multicolumn{1}{|c|}{}                               &                                   & \textbf{(b)}                       & \textbf{0.758}              & \textbf{0.763}              & \textbf{0.625}              \\ \cline{3-6} 
\multicolumn{1}{|c|}{}                               &                                   & \textbf{(c)}                       & 0.747                       & 0.754                       & 0.568                       \\ \cline{2-6} 
\multicolumn{1}{|c|}{}                               & \multirow{3}{*}{\textbf{350-100}} & \textbf{(a)}                       & 0.489                       & 0.500                       & 0.354                       \\ \cline{3-6} 
\multicolumn{1}{|c|}{}                               &                                   & \textbf{(b)}                       & 0.749                       & 0.759                       & 0.624                       \\ \cline{3-6} 
\multicolumn{1}{|c|}{}                               &                                   & \textbf{(c)}                       & 0.737                       & 0.742                       & 0.570                       \\ \hline \hline
\multicolumn{1}{|c|}{\multirow{3}{*}{\textbf{300}}}  & \multirow{3}{*}{\textbf{350-50}}  & \textbf{(a)}                       & 0.498                       & 0.509                       & 0.360                       \\ \cline{3-6} 
\multicolumn{1}{|c|}{}                               &                                   & \textbf{(b)}                       & \textbf{0.764}              & \textbf{0.771}              & \textbf{0.623}              \\ \cline{3-6} 
\multicolumn{1}{|c|}{}                               &                                   & \textbf{(c)}                       & 0.736                       & 0.742                       & 0.588                       \\ \hline
\end{tabular}}
\label{tab:tab04}
\vspace{0.10cm}
\scriptsize{\\
Note: the bold entries represent the highest PCC value obtained using the neural decoders in x, y and z directions for each window size.}
\end{table}

\subsection{Inter-subject decoding analysis}\label{sec:sub_ind}
In this analysis, the spectral features in FB1 frequency band is considered for similar reason as mentioned in section \ref{sec:spectral}. The decoding performance of each neural decoder for different EEG lags and test subject is depicted in Table \ref{tab:tab05} as PCC value. The mean PCC value of all the test subjects is calculated for each EEG lag and direction to evaluate the performance of neural decoders. It may be observed that decoding performance is improved with the increment in the size of EEG lag window. 
\begin{table*}[t]
\caption{Inter-Subject analysis of (a) mLR, (b) MLP, and (c) CNN-LSTM neural decoders with time lags of 150 ms, 200 ms, 250 ms, 300 ms and 350 ms.}
\scalebox{0.81}{
\centering
\begin{tabular}{|c|c|c|cccccccccccc|c|}
\hline
\multirow{2}{*}{\textbf{EEG Lags}} & \multirow{2}{*}{\textbf{Decoders}} & \multirow{2}{*}{\textbf{Direction}} & \multicolumn{12}{c|}{\textbf{TEST SUBJECT}}                                                                                                                                                                                                                                                                                                                                           &                  \\ \cline{4-16} 
                                    &                                    &                                     & \multicolumn{1}{c|}{\textbf{S12}} & \multicolumn{1}{c|}{\textbf{S11}} & \multicolumn{1}{c|}{\textbf{S10}} & \multicolumn{1}{c|}{\textbf{S09}} & \multicolumn{1}{c|}{\textbf{S08}} & \multicolumn{1}{c|}{\textbf{S07}} & \multicolumn{1}{c|}{\textbf{S06}} & \multicolumn{1}{c|}{\textbf{S05}} & \multicolumn{1}{c|}{\textbf{S04}} & \multicolumn{1}{c|}{\textbf{S03}} & \multicolumn{1}{c|}{\textbf{S02}} & \textbf{S01} & \textbf{Average} \\ \hline \hline
\multirow{9}{*}{\textbf{150}}       & \multirow{3}{*}{\textbf{(a)}}      & \textbf{x}                          & \multicolumn{1}{c|}{0.123}        & \multicolumn{1}{c|}{0.153}        & \multicolumn{1}{c|}{0.389}        & \multicolumn{1}{c|}{0.312}        & \multicolumn{1}{c|}{0.178}        & \multicolumn{1}{c|}{0.248}        & \multicolumn{1}{c|}{0.305}        & \multicolumn{1}{c|}{0.297}        & \multicolumn{1}{c|}{0.401}        & \multicolumn{1}{c|}{0.296}        & \multicolumn{1}{c|}{0.177}        & 0.431        & 0.276            \\ \cline{3-16} 
                                    &                                    & \textbf{y}                          & \multicolumn{1}{c|}{0.144}        & \multicolumn{1}{c|}{0.159}        & \multicolumn{1}{c|}{0.399}        & \multicolumn{1}{c|}{0.314}        & \multicolumn{1}{c|}{0.178}        & \multicolumn{1}{c|}{0.246}        & \multicolumn{1}{c|}{0.289}        & \multicolumn{1}{c|}{0.284}        & \multicolumn{1}{c|}{0.413}        & \multicolumn{1}{c|}{0.300}        & \multicolumn{1}{c|}{0.181}        & 0.434        & 0.278            \\ \cline{3-16} 
                                    &                                    & \textbf{z}                          & \multicolumn{1}{c|}{0.287}        & \multicolumn{1}{c|}{0.179}        & \multicolumn{1}{c|}{0.224}        & \multicolumn{1}{c|}{0.199}        & \multicolumn{1}{c|}{0.254}        & \multicolumn{1}{c|}{0.257}        & \multicolumn{1}{c|}{0.138}        & \multicolumn{1}{c|}{0.293}        & \multicolumn{1}{c|}{0.334}        & \multicolumn{1}{c|}{0.306}        & \multicolumn{1}{c|}{0.115}        & 0.309        & 0.241            \\ \cline{2-16} 
                                    & \multirow{3}{*}{\textbf{(b)}}      & \textbf{x}                          & \multicolumn{1}{c|}{0.651}        & \multicolumn{1}{c|}{0.687}        & \multicolumn{1}{c|}{0.779}        & \multicolumn{1}{c|}{0.768}        & \multicolumn{1}{c|}{0.764}        & \multicolumn{1}{c|}{0.743}        & \multicolumn{1}{c|}{0.779}        & \multicolumn{1}{c|}{0.774}        & \multicolumn{1}{c|}{0.735}        & \multicolumn{1}{c|}{0.685}        & \multicolumn{1}{c|}{0.684}        & 0.681        & 0.727            \\ \cline{3-16} 
                                    &                                    & \textbf{y}                          & \multicolumn{1}{c|}{0.688}        & \multicolumn{1}{c|}{0.709}        & \multicolumn{1}{c|}{0.785}        & \multicolumn{1}{c|}{0.787}        & \multicolumn{1}{c|}{0.768}        & \multicolumn{1}{c|}{0.755}        & \multicolumn{1}{c|}{0.795}        & \multicolumn{1}{c|}{0.793}        & \multicolumn{1}{c|}{0.748}        & \multicolumn{1}{c|}{0.705}        & \multicolumn{1}{c|}{0.697}        & 0.719        & 0.746            \\ \cline{3-16} 
                                    &                                    & \textbf{z}                          & \multicolumn{1}{c|}{0.429}        & \multicolumn{1}{c|}{0.362}        & \multicolumn{1}{c|}{0.635}        & \multicolumn{1}{c|}{0.461}        & \multicolumn{1}{c|}{0.666}        & \multicolumn{1}{c|}{0.428}        & \multicolumn{1}{c|}{0.526}        & \multicolumn{1}{c|}{0.604}        & \multicolumn{1}{c|}{0.644}        & \multicolumn{1}{c|}{0.702}        & \multicolumn{1}{c|}{0.359}        & 0.586        & \textbf{0.533}   \\ \cline{2-16} 
                                    & \multirow{3}{*}{\textbf{(c)}}      & \textbf{x}                          & \multicolumn{1}{c|}{0.635}        & \multicolumn{1}{c|}{0.696}        & \multicolumn{1}{c|}{0.791}        & \multicolumn{1}{c|}{0.762}        & \multicolumn{1}{c|}{0.767}        & \multicolumn{1}{c|}{0.752}        & \multicolumn{1}{c|}{0.788}        & \multicolumn{1}{c|}{0.775}        & \multicolumn{1}{c|}{0.731}        & \multicolumn{1}{c|}{0.707}        & \multicolumn{1}{c|}{0.703}        & 0.655        & \textbf{0.730}   \\ \cline{3-16} 
                                    &                                    & \textbf{y}                          & \multicolumn{1}{c|}{0.671}        & \multicolumn{1}{c|}{0.716}        & \multicolumn{1}{c|}{0.795}        & \multicolumn{1}{c|}{0.783}        & \multicolumn{1}{c|}{0.773}        & \multicolumn{1}{c|}{0.761}        & \multicolumn{1}{c|}{0.802}        & \multicolumn{1}{c|}{0.789}        & \multicolumn{1}{c|}{0.738}        & \multicolumn{1}{c|}{0.727}        & \multicolumn{1}{c|}{0.713}        & 0.697        & \textbf{0.747}   \\ \cline{3-16} 
                                    &                                    & \textbf{z}                          & \multicolumn{1}{c|}{0.405}        & \multicolumn{1}{c|}{0.345}        & \multicolumn{1}{c|}{0.594}        & \multicolumn{1}{c|}{0.440}        & \multicolumn{1}{c|}{0.632}        & \multicolumn{1}{c|}{0.400}        & \multicolumn{1}{c|}{0.507}        & \multicolumn{1}{c|}{0.538}        & \multicolumn{1}{c|}{0.592}        & \multicolumn{1}{c|}{0.667}        & \multicolumn{1}{c|}{0.335}        & 0.536        & 0.499            \\ \hline \hline
\multirow{9}{*}{\textbf{200}}       & \multirow{3}{*}{\textbf{(a)}}      & \textbf{x}                          & \multicolumn{1}{c|}{0.106}        & \multicolumn{1}{c|}{0.099}        & \multicolumn{1}{c|}{0.136}        & \multicolumn{1}{c|}{0.325}        & \multicolumn{1}{c|}{0.170}        & \multicolumn{1}{c|}{0.267}        & \multicolumn{1}{c|}{0.275}        & \multicolumn{1}{c|}{0.301}        & \multicolumn{1}{c|}{0.364}        & \multicolumn{1}{c|}{0.288}        & \multicolumn{1}{c|}{0.211}        & 0.423        & 0.247            \\ \cline{3-16} 
                                    &                                    & \textbf{y}                          & \multicolumn{1}{c|}{0.120}        & \multicolumn{1}{c|}{0.106}        & \multicolumn{1}{c|}{0.123}        & \multicolumn{1}{c|}{0.327}        & \multicolumn{1}{c|}{0.171}        & \multicolumn{1}{c|}{0.261}        & \multicolumn{1}{c|}{0.260}        & \multicolumn{1}{c|}{0.291}        & \multicolumn{1}{c|}{0.379}        & \multicolumn{1}{c|}{0.295}        & \multicolumn{1}{c|}{0.213}        & 0.415        & 0.247            \\ \cline{3-16} 
                                    &                                    & \textbf{z}                          & \multicolumn{1}{c|}{0.302}        & \multicolumn{1}{c|}{0.186}        & \multicolumn{1}{c|}{0.210}        & \multicolumn{1}{c|}{0.212}        & \multicolumn{1}{c|}{0.235}        & \multicolumn{1}{c|}{0.274}        & \multicolumn{1}{c|}{0.151}        & \multicolumn{1}{c|}{0.289}        & \multicolumn{1}{c|}{0.349}        & \multicolumn{1}{c|}{0.269}        & \multicolumn{1}{c|}{0.198}        & 0.307        & 0.249            \\ \cline{2-16} 
                                    & \multirow{3}{*}{\textbf{(b)}}      & \textbf{x}                          & \multicolumn{1}{c|}{0.655}        & \multicolumn{1}{c|}{0.660}        & \multicolumn{1}{c|}{0.745}        & \multicolumn{1}{c|}{0.789}        & \multicolumn{1}{c|}{0.765}        & \multicolumn{1}{c|}{0.755}        & \multicolumn{1}{c|}{0.804}        & \multicolumn{1}{c|}{0.779}        & \multicolumn{1}{c|}{0.723}        & \multicolumn{1}{c|}{0.720}        & \multicolumn{1}{c|}{0.692}        & 0.669        & 0.730            \\ \cline{3-16} 
                                    &                                    & \textbf{y}                          & \multicolumn{1}{c|}{0.697}        & \multicolumn{1}{c|}{0.695}        & \multicolumn{1}{c|}{0.750}        & \multicolumn{1}{c|}{0.813}        & \multicolumn{1}{c|}{0.775}        & \multicolumn{1}{c|}{0.765}        & \multicolumn{1}{c|}{0.817}        & \multicolumn{1}{c|}{0.798}        & \multicolumn{1}{c|}{0.733}        & \multicolumn{1}{c|}{0.742}        & \multicolumn{1}{c|}{0.701}        & 0.708        & 0.750            \\ \cline{3-16} 
                                    &                                    & \textbf{z}                          & \multicolumn{1}{c|}{0.465}        & \multicolumn{1}{c|}{0.373}        & \multicolumn{1}{c|}{0.603}        & \multicolumn{1}{c|}{0.483}        & \multicolumn{1}{c|}{0.674}        & \multicolumn{1}{c|}{0.432}        & \multicolumn{1}{c|}{0.549}        & \multicolumn{1}{c|}{0.591}        & \multicolumn{1}{c|}{0.645}        & \multicolumn{1}{c|}{0.711}        & \multicolumn{1}{c|}{0.412}        & 0.575        & \textbf{0.543}   \\ \cline{2-16} 
                                    & \multirow{3}{*}{\textbf{(c)}}      & \textbf{x}                          & \multicolumn{1}{c|}{0.657}        & \multicolumn{1}{c|}{0.672}        & \multicolumn{1}{c|}{0.769}        & \multicolumn{1}{c|}{0.778}        & \multicolumn{1}{c|}{0.786}        & \multicolumn{1}{c|}{0.758}        & \multicolumn{1}{c|}{0.801}        & \multicolumn{1}{c|}{0.790}        & \multicolumn{1}{c|}{0.736}        & \multicolumn{1}{c|}{0.723}        & \multicolumn{1}{c|}{0.722}        & 0.680        & \textbf{0.739}   \\ \cline{3-16} 
                                    &                                    & \textbf{y}                          & \multicolumn{1}{c|}{0.693}        & \multicolumn{1}{c|}{0.699}        & \multicolumn{1}{c|}{0.776}        & \multicolumn{1}{c|}{0.802}        & \multicolumn{1}{c|}{0.796}        & \multicolumn{1}{c|}{0.767}        & \multicolumn{1}{c|}{0.818}        & \multicolumn{1}{c|}{0.808}        & \multicolumn{1}{c|}{0.750}        & \multicolumn{1}{c|}{0.741}        & \multicolumn{1}{c|}{0.735}        & 0.720        & \textbf{0.759}   \\ \cline{3-16} 
                                    &                                    & \textbf{z}                          & \multicolumn{1}{c|}{0.439}        & \multicolumn{1}{c|}{0.331}        & \multicolumn{1}{c|}{0.554}        & \multicolumn{1}{c|}{0.441}        & \multicolumn{1}{c|}{0.633}        & \multicolumn{1}{c|}{0.396}        & \multicolumn{1}{c|}{0.512}        & \multicolumn{1}{c|}{0.561}        & \multicolumn{1}{c|}{0.586}        & \multicolumn{1}{c|}{0.656}        & \multicolumn{1}{c|}{0.385}        & 0.532        & 0.502            \\ \hline \hline
\multirow{9}{*}{\textbf{250}}       & \multirow{3}{*}{\textbf{(a)}}      & \textbf{x}                          & \multicolumn{1}{c|}{0.100}        & \multicolumn{1}{c|}{0.072}        & \multicolumn{1}{c|}{0.099}        & \multicolumn{1}{c|}{0.305}        & \multicolumn{1}{c|}{0.143}        & \multicolumn{1}{c|}{0.262}        & \multicolumn{1}{c|}{0.261}        & \multicolumn{1}{c|}{0.286}        & \multicolumn{1}{c|}{0.360}        & \multicolumn{1}{c|}{0.289}        & \multicolumn{1}{c|}{0.170}        & 0.418        & 0.230            \\ \cline{3-16} 
                                    &                                    & \textbf{y}                          & \multicolumn{1}{c|}{0.106}        & \multicolumn{1}{c|}{0.078}        & \multicolumn{1}{c|}{0.084}        & \multicolumn{1}{c|}{0.308}        & \multicolumn{1}{c|}{0.143}        & \multicolumn{1}{c|}{0.246}        & \multicolumn{1}{c|}{0.243}        & \multicolumn{1}{c|}{0.281}        & \multicolumn{1}{c|}{0.374}        & \multicolumn{1}{c|}{0.293}        & \multicolumn{1}{c|}{0.175}        & 0.406        & 0.228            \\ \cline{3-16} 
                                    &                                    & \textbf{z}                          & \multicolumn{1}{c|}{0.318}        & \multicolumn{1}{c|}{0.143}        & \multicolumn{1}{c|}{0.187}        & \multicolumn{1}{c|}{0.232}        & \multicolumn{1}{c|}{0.224}        & \multicolumn{1}{c|}{0.286}        & \multicolumn{1}{c|}{0.159}        & \multicolumn{1}{c|}{0.266}        & \multicolumn{1}{c|}{0.356}        & \multicolumn{1}{c|}{0.255}        & \multicolumn{1}{c|}{0.217}        & 0.295        & 0.245            \\ \cline{2-16} 
                                    & \multirow{3}{*}{\textbf{(b)}}      & \textbf{x}                          & \multicolumn{1}{c|}{0.644}        & \multicolumn{1}{c|}{0.693}        & \multicolumn{1}{c|}{0.750}        & \multicolumn{1}{c|}{0.771}        & \multicolumn{1}{c|}{0.777}        & \multicolumn{1}{c|}{0.764}        & \multicolumn{1}{c|}{0.794}        & \multicolumn{1}{c|}{0.761}        & \multicolumn{1}{c|}{0.744}        & \multicolumn{1}{c|}{0.712}        & \multicolumn{1}{c|}{0.697}        & 0.677        & 0.732            \\ \cline{3-16} 
                                    &                                    & \textbf{y}                          & \multicolumn{1}{c|}{0.690}        & \multicolumn{1}{c|}{0.715}        & \multicolumn{1}{c|}{0.761}        & \multicolumn{1}{c|}{0.796}        & \multicolumn{1}{c|}{0.782}        & \multicolumn{1}{c|}{0.775}        & \multicolumn{1}{c|}{0.805}        & \multicolumn{1}{c|}{0.788}        & \multicolumn{1}{c|}{0.754}        & \multicolumn{1}{c|}{0.734}        & \multicolumn{1}{c|}{0.711}        & 0.714        & 0.752            \\ \cline{3-16} 
                                    &                                    & \textbf{z}                          & \multicolumn{1}{c|}{0.484}        & \multicolumn{1}{c|}{0.349}        & \multicolumn{1}{c|}{0.627}        & \multicolumn{1}{c|}{0.486}        & \multicolumn{1}{c|}{0.674}        & \multicolumn{1}{c|}{0.451}        & \multicolumn{1}{c|}{0.547}        & \multicolumn{1}{c|}{0.609}        & \multicolumn{1}{c|}{0.661}        & \multicolumn{1}{c|}{0.712}        & \multicolumn{1}{c|}{0.405}        & 0.599        & \textbf{0.550}   \\ \cline{2-16} 
                                    & \multirow{3}{*}{\textbf{(c)}}      & \textbf{x}                          & \multicolumn{1}{c|}{0.684}        & \multicolumn{1}{c|}{0.676}        & \multicolumn{1}{c|}{0.773}        & \multicolumn{1}{c|}{0.764}        & \multicolumn{1}{c|}{0.789}        & \multicolumn{1}{c|}{0.768}        & \multicolumn{1}{c|}{0.801}        & \multicolumn{1}{c|}{0.793}        & \multicolumn{1}{c|}{0.748}        & \multicolumn{1}{c|}{0.712}        & \multicolumn{1}{c|}{0.716}        & 0.669        & \textbf{0.741}   \\ \cline{3-16} 
                                    &                                    & \textbf{y}                          & \multicolumn{1}{c|}{0.722}        & \multicolumn{1}{c|}{0.703}        & \multicolumn{1}{c|}{0.780}        & \multicolumn{1}{c|}{0.785}        & \multicolumn{1}{c|}{0.797}        & \multicolumn{1}{c|}{0.779}        & \multicolumn{1}{c|}{0.813}        & \multicolumn{1}{c|}{0.812}        & \multicolumn{1}{c|}{0.759}        & \multicolumn{1}{c|}{0.729}        & \multicolumn{1}{c|}{0.724}        & 0.705        & \textbf{0.759}   \\ \cline{3-16} 
                                    &                                    & \textbf{z}                          & \multicolumn{1}{c|}{0.455}        & \multicolumn{1}{c|}{0.298}        & \multicolumn{1}{c|}{0.545}        & \multicolumn{1}{c|}{0.443}        & \multicolumn{1}{c|}{0.632}        & \multicolumn{1}{c|}{0.403}        & \multicolumn{1}{c|}{0.482}        & \multicolumn{1}{c|}{0.538}        & \multicolumn{1}{c|}{0.570}        & \multicolumn{1}{c|}{0.665}        & \multicolumn{1}{c|}{0.385}        & 0.551        & 0.497            \\ \hline \hline
\multirow{9}{*}{\textbf{300}}       & \multirow{3}{*}{\textbf{(a)}}      & \textbf{x}                          & \multicolumn{1}{c|}{0.091}        & \multicolumn{1}{c|}{0.069}        & \multicolumn{1}{c|}{0.132}        & \multicolumn{1}{c|}{0.275}        & \multicolumn{1}{c|}{0.134}        & \multicolumn{1}{c|}{0.235}        & \multicolumn{1}{c|}{0.346}        & \multicolumn{1}{c|}{0.270}        & \multicolumn{1}{c|}{0.357}        & \multicolumn{1}{c|}{0.262}        & \multicolumn{1}{c|}{0.133}        & 0.439        & 0.229            \\ \cline{3-16} 
                                    &                                    & \textbf{y}                          & \multicolumn{1}{c|}{0.094}        & \multicolumn{1}{c|}{0.073}        & \multicolumn{1}{c|}{0.121}        & \multicolumn{1}{c|}{0.276}        & \multicolumn{1}{c|}{0.130}        & \multicolumn{1}{c|}{0.218}        & \multicolumn{1}{c|}{0.340}        & \multicolumn{1}{c|}{0.270}        & \multicolumn{1}{c|}{0.380}        & \multicolumn{1}{c|}{0.265}        & \multicolumn{1}{c|}{0.141}        & 0.431        & 0.228            \\ \cline{3-16} 
                                    &                                    & \textbf{z}                          & \multicolumn{1}{c|}{0.317}        & \multicolumn{1}{c|}{0.171}        & \multicolumn{1}{c|}{0.196}        & \multicolumn{1}{c|}{0.250}        & \multicolumn{1}{c|}{0.224}        & \multicolumn{1}{c|}{0.272}        & \multicolumn{1}{c|}{0.171}        & \multicolumn{1}{c|}{0.226}        & \multicolumn{1}{c|}{0.341}        & \multicolumn{1}{c|}{0.211}        & \multicolumn{1}{c|}{0.135}        & 0.272        & 0.232            \\ \cline{2-16} 
                                    & \multirow{3}{*}{\textbf{(b)}}      & \textbf{x}                          & \multicolumn{1}{c|}{0.635}        & \multicolumn{1}{c|}{0.660}        & \multicolumn{1}{c|}{0.755}        & \multicolumn{1}{c|}{0.769}        & \multicolumn{1}{c|}{0.765}        & \multicolumn{1}{c|}{0.759}        & \multicolumn{1}{c|}{0.822}        & \multicolumn{1}{c|}{0.763}        & \multicolumn{1}{c|}{0.728}        & \multicolumn{1}{c|}{0.705}        & \multicolumn{1}{c|}{0.679}        & 0.692        & 0.728            \\ \cline{3-16} 
                                    &                                    & \textbf{y}                          & \multicolumn{1}{c|}{0.683}        & \multicolumn{1}{c|}{0.682}        & \multicolumn{1}{c|}{0.766}        & \multicolumn{1}{c|}{0.785}        & \multicolumn{1}{c|}{0.774}        & \multicolumn{1}{c|}{0.770}        & \multicolumn{1}{c|}{0.835}        & \multicolumn{1}{c|}{0.788}        & \multicolumn{1}{c|}{0.738}        & \multicolumn{1}{c|}{0.728}        & \multicolumn{1}{c|}{0.697}        & 0.728        & 0.748            \\ \cline{3-16} 
                                    &                                    & \textbf{z}                          & \multicolumn{1}{c|}{0.478}        & \multicolumn{1}{c|}{0.369}        & \multicolumn{1}{c|}{0.618}        & \multicolumn{1}{c|}{0.499}        & \multicolumn{1}{c|}{0.687}        & \multicolumn{1}{c|}{0.452}        & \multicolumn{1}{c|}{0.592}        & \multicolumn{1}{c|}{0.576}        & \multicolumn{1}{c|}{0.642}        & \multicolumn{1}{c|}{0.710}        & \multicolumn{1}{c|}{0.368}        & 0.620        & \textbf{0.551}   \\ \cline{2-16} 
                                    & \multirow{3}{*}{\textbf{(c)}}      & \textbf{x}                          & \multicolumn{1}{c|}{0.655}        & \multicolumn{1}{c|}{0.697}        & \multicolumn{1}{c|}{0.779}        & \multicolumn{1}{c|}{0.772}        & \multicolumn{1}{c|}{0.787}        & \multicolumn{1}{c|}{0.776}        & \multicolumn{1}{c|}{0.800}        & \multicolumn{1}{c|}{0.798}        & \multicolumn{1}{c|}{0.734}        & \multicolumn{1}{c|}{0.730}        & \multicolumn{1}{c|}{0.713}        & 0.708        & \textbf{0.746}   \\ \cline{3-16} 
                                    &                                    & \textbf{y}                          & \multicolumn{1}{c|}{0.694}        & \multicolumn{1}{c|}{0.719}        & \multicolumn{1}{c|}{0.788}        & \multicolumn{1}{c|}{0.795}        & \multicolumn{1}{c|}{0.798}        & \multicolumn{1}{c|}{0.791}        & \multicolumn{1}{c|}{0.819}        & \multicolumn{1}{c|}{0.820}        & \multicolumn{1}{c|}{0.745}        & \multicolumn{1}{c|}{0.749}        & \multicolumn{1}{c|}{0.733}        & 0.743        & \textbf{0.766}   \\ \cline{3-16} 
                                    &                                    & \textbf{z}                          & \multicolumn{1}{c|}{0.414}        & \multicolumn{1}{c|}{0.307}        & \multicolumn{1}{c|}{0.525}        & \multicolumn{1}{c|}{0.430}        & \multicolumn{1}{c|}{0.607}        & \multicolumn{1}{c|}{0.397}        & \multicolumn{1}{c|}{0.513}        & \multicolumn{1}{c|}{0.498}        & \multicolumn{1}{c|}{0.532}        & \multicolumn{1}{c|}{0.628}        & \multicolumn{1}{c|}{0.325}        & 0.574        & 0.479            \\ \hline \hline
\multirow{9}{*}{\textbf{350}}       & \multirow{3}{*}{\textbf{(a)}}      & \textbf{x}                          & \multicolumn{1}{c|}{0.080}        & \multicolumn{1}{c|}{0.076}        & \multicolumn{1}{c|}{0.134}        & \multicolumn{1}{c|}{0.265}        & \multicolumn{1}{c|}{0.137}        & \multicolumn{1}{c|}{0.248}        & \multicolumn{1}{c|}{0.258}        & \multicolumn{1}{c|}{0.272}        & \multicolumn{1}{c|}{0.339}        & \multicolumn{1}{c|}{0.301}        & \multicolumn{1}{c|}{0.143}        & 0.463        & 0.226            \\ \cline{3-16} 
                                    &                                    & \textbf{y}                          & \multicolumn{1}{c|}{0.079}        & \multicolumn{1}{c|}{0.075}        & \multicolumn{1}{c|}{0.119}        & \multicolumn{1}{c|}{0.264}        & \multicolumn{1}{c|}{0.136}        & \multicolumn{1}{c|}{0.228}        & \multicolumn{1}{c|}{0.244}        & \multicolumn{1}{c|}{0.273}        & \multicolumn{1}{c|}{0.363}        & \multicolumn{1}{c|}{0.299}        & \multicolumn{1}{c|}{0.155}        & 0.452        & 0.224            \\ \cline{3-16} 
                                    &                                    & \textbf{z}                          & \multicolumn{1}{c|}{0.317}        & \multicolumn{1}{c|}{0.149}        & \multicolumn{1}{c|}{0.167}        & \multicolumn{1}{c|}{0.264}        & \multicolumn{1}{c|}{0.200}        & \multicolumn{1}{c|}{0.260}        & \multicolumn{1}{c|}{0.177}        & \multicolumn{1}{c|}{0.207}        & \multicolumn{1}{c|}{0.343}        & \multicolumn{1}{c|}{0.205}        & \multicolumn{1}{c|}{0.160}        & 0.248        & 0.225            \\ \cline{2-16} 
                                    & \multirow{3}{*}{\textbf{(b)}}      & \textbf{x}                          & \multicolumn{1}{c|}{0.631}        & \multicolumn{1}{c|}{0.692}        & \multicolumn{1}{c|}{0.767}        & \multicolumn{1}{c|}{0.765}        & \multicolumn{1}{c|}{0.759}        & \multicolumn{1}{c|}{0.754}        & \multicolumn{1}{c|}{0.806}        & \multicolumn{1}{c|}{0.775}        & \multicolumn{1}{c|}{0.716}        & \multicolumn{1}{c|}{0.713}        & \multicolumn{1}{c|}{0.707}        & 0.686        & 0.731            \\ \cline{3-16} 
                                    &                                    & \textbf{y}                          & \multicolumn{1}{c|}{0.673}        & \multicolumn{1}{c|}{0.710}        & \multicolumn{1}{c|}{0.776}        & \multicolumn{1}{c|}{0.788}        & \multicolumn{1}{c|}{0.767}        & \multicolumn{1}{c|}{0.765}        & \multicolumn{1}{c|}{0.812}        & \multicolumn{1}{c|}{0.800}        & \multicolumn{1}{c|}{0.727}        & \multicolumn{1}{c|}{0.729}        & \multicolumn{1}{c|}{0.725}        & 0.728        & 0.750            \\ \cline{3-16} 
                                    &                                    & \textbf{z}                          & \multicolumn{1}{c|}{0.495}        & \multicolumn{1}{c|}{0.378}        & \multicolumn{1}{c|}{0.617}        & \multicolumn{1}{c|}{0.497}        & \multicolumn{1}{c|}{0.676}        & \multicolumn{1}{c|}{0.459}        & \multicolumn{1}{c|}{0.569}        & \multicolumn{1}{c|}{0.591}        & \multicolumn{1}{c|}{0.636}        & \multicolumn{1}{c|}{0.700}        & \multicolumn{1}{c|}{0.400}        & 0.611        & \textbf{0.552}   \\ \cline{2-16} 
                                    &         \multirow{3}{*}{\textbf{(c)}}      & \textbf{x}                          & \multicolumn{1}{c|}{0.690}        & \multicolumn{1}{c|}{0.712}        & \multicolumn{1}{c|}{0.798}        & \multicolumn{1}{c|}{0.776}        & \multicolumn{1}{c|}{0.799}        & \multicolumn{1}{c|}{0.787}        & \multicolumn{1}{c|}{0.810}        & \multicolumn{1}{c|}{0.808}        & \multicolumn{1}{c|}{0.747}        & \multicolumn{1}{c|}{0.737}        & \multicolumn{1}{c|}{0.726}        & 0.728        & \textbf{0.760}   \\ \cline{3-16} 
                                    &                                    & \textbf{y}                          & \multicolumn{1}{c|}{0.728}        & \multicolumn{1}{c|}{0.732}        & \multicolumn{1}{c|}{0.807}        & \multicolumn{1}{c|}{0.799}        & \multicolumn{1}{c|}{0.805}        & \multicolumn{1}{c|}{0.790}        & \multicolumn{1}{c|}{0.829}        & \multicolumn{1}{c|}{0.829}        & \multicolumn{1}{c|}{0.753}        & \multicolumn{1}{c|}{0.753}        & \multicolumn{1}{c|}{0.746}        & 0.764        & \textbf{0.778}   \\ \cline{3-16} 
                                    &                                    & \textbf{z}                          & \multicolumn{1}{c|}{0.462}        & \multicolumn{1}{c|}{0.313}        & \multicolumn{1}{c|}{0.519}        & \multicolumn{1}{c|}{0.434}        & \multicolumn{1}{c|}{0.617}        & \multicolumn{1}{c|}{0.385}        & \multicolumn{1}{c|}{0.506}        & \multicolumn{1}{c|}{0.487}        & \multicolumn{1}{c|}{0.498}        & \multicolumn{1}{c|}{0.620}        & \multicolumn{1}{c|}{0.365}        & 0.565        & 0.481            \\ \hline
\end{tabular}}
\label{tab:tab05}
\vspace{0.10cm}
\centering
\scriptsize{\\
Note: the bold entries represent the highest mean PCC value obtained using the neural decoders in x, y and z directions for each EEG lag.}
\end{table*}

\section{Discussion}\label{sec:discussion}

In this study, the affect of spectral features on the hand trajectory decoding performance is highlighted. The EEG data with spectral features in delta band, FB1, has exhibited significantly better decoding performance for grasp and lift task. Spectral features from FB1 frequency band has shown best decoding performance in the x and y-directions, while the performance is similar to FB7 frequency band in the z-direction. It is evident from the literature that low frequency spectral features have major contribution in decoding hand movements\cite{chouhan2018wavlet,hosseini2022continuous}. In particular, low frequency EEG signals have been utilized for efficient motor decoding\cite{ofner2017upper, Sosnik2020,jain2022premovnet}. In \cite{korik2018decoding}, spectral features in alpha, beta and low gamma frequency bands have been utilized to decode 3-D hand trajectory using EEG signals. In this work, a comparative analysis is presented for all the frequency bands and combinations. It may be observed that proposed deep learning-based neural decoders with EEG spectral features in theta, alpha and beta frequency bands have shown decent decoding performance (PCC $>$ 0.5 in x and y-directions while PCC $>$ 0.4 in z-direction, refer Table \ref{tab:tab03}), but poor decoding performance in low-gamma frequency band for grasp and lift task. However, the neural decoders with spectral features in the delta band have shown superior performance. Hence, FB1 spectral features is considered for the analysis in further section \ref{sec:eeg_lag} and \ref{sec:sub_ind}.

3-D hand trajectory decoding using different lag windows is investigated in section \ref{sec:eeg_lag}. The lag windows with pre-movement EEG data and spectral features in delta band have been utilized for decoding 3-D hand trajectory for reach and grasp task. The pre-movement EEG data, with at-least 50 ms prior to the movement onset, has been utilized for hand trajectory decoding. Table \ref{tab:tab04} shows the correlation values for neural decoders with distinct lag windows. In particular, MLP based neural decoder has outperformed mLR and CNN-LSTM based models for each lag window size. It may be observed that the decoding performance has improved marginally with an increase in the EEG lag window size. The motor intention detection using pre-movement EEG data could be helpful in BCI applications such as controlling prosthesis and power-augmented exosuit. This is inline with the fact that motor-movement information is present around 300 msec prior to the actual movement\cite{source2022}.

Inter-subject BCI system for robust decoding performance is investigated in section \ref{sec:sub_ind}. In particular, LOSO cross-validation approach is adopted to evaluate inter-subject decoding performance of the proposed neural decoders. EEG signals with spectral features in delta band and EEG lags are used for 3-D hand trajectory decoding during grasp and lift task. Table \ref{tab:tab05} shows the correlation coefficient values for inter-subject analysis with various EEG lags. For each EEG lag, performance of the neural decoders is evaluated in LOSO manner. The results are shown for each test subject and mean PCC values are highlighted. For each EEG lag, efficient inter-subject learning performance of CNN-LSTM based neural decoder may be observed in x and y-directions while MLP-based neural decoders have shown better decoding performance in z-direction. This analysis shows the feasibility of inter-subject BCI system to decode 3-D hand trajectory during grasp and lift task. The proposed neural decoders have shown relatively low performance to decode hand trajectory in z-direction. Therefore, better neural features in conjunction with optimized neural decoder need to be explored for improving the decoding performance in z-direction.

\section{Conclusion}\label{sec:conclusion}            

In this study, deep learning-based neural decoders have been proposed for efficient 3-D hand kinematics decoding using pre-movement EEG signals. In particular, MLP and CNN-LSTM based models are proposed. The pre-movement neural information encoded in EEG signals has been utilized using EEG lags for efficient hand trajectory decoding. Spectral features for efficient neural decoding for 3-D hand kinematics has been investigated. Additionally, the feasibility of inter-subject hand trajectory decoding has been examined. PCC analysis has been adopted to establish the effectiveness of proposed neural decoders. The results show that the proposed neural decoders could decode the 3D hand trajectory for inter-subject BCI applications. This study provides the feasibility of inter-subject BCI system that can expedite real-time control of rehabilitation devices and prosthesis. 

\section*{Acknowledgment}
This research work was supported in part by DRDO - JATC project with project number RP04191G. The authors would like to thank Prof. Sitikantha Roy, Prof. Shubhendu Bhasin, and Prof. Sushma Santapuri from Indian Institute of Technology Delhi (IITD), and Dr. Suriya Prakash from All India Institute of Medical Sciences (AIIMS) Delhi for their discussion and constructive comments during the preparation of the manuscript.

\bibliographystyle{IEEEtran}

\bibliography{Draft_v01_cybernetics}




\end{document}